\documentclass[12pt]{article}

\title{Pressure Correction in Classical Density Functional Theory: Hyper Netted Chain and Hard Sphere Bridge Functionals.}
\author{
  Volodymyr Sergiievskyi,
  Guillaume Jeanmairet, 
  Maximilien Levesque, 
  Daniel Borgis}

\usepackage{amsmath}
\usepackage{amssymb}

\usepackage[utf8]{inputenc}

\usepackage{morefloats}
\usepackage[top=2cm,left=2cm,right=2cm,bottom=2cm]{geometry}
\usepackage{graphicx}
\usepackage{enumitem}
\usepackage[usenames]{color}

\usepackage{verbatim}

\newcommand{\Angstr}{ \ensuremath{ \text{\AA} } }

\newcommand{\colorcomment}[2] { {\color{#1} \textbf{#2} } }

\newcommand{\rcomment}[1]{ \colorcomment{red}{#1}  }
\newcommand{\bcomment}[1]{ \colorcomment{blue}{#1} }

\newcommand{\maxcomment}[1]{  \rcomment{Max: #1} }
\newcommand{\voovcomment}[1]{  \bcomment{voovrat: #1} }

\renewcommand{\maxcomment}[1] {}
\renewcommand{\voovcomment}[1] {}

\newcommand{\subtopic}[1]{}

\begin{document}

\maketitle

\begin{abstract}

Low accuracy of the Solvation Free Energy (SFE) calculation is a known problem of the numerical methods of the Integral Equation Theory of Liquids and the Classical Density Functional Theory (Classical DFT).
Although functionals with empirical corrections can essentially improve the predictability of the methods, their universality is still a question.
In our recent paper we connected the SFE calculation errors with the incorrect pressure in the Classical DFT and proposed the a posteriory correction to improve the results (J. Phys. Chem. Lett., 5, 1925-1942 ).
This paper raised a discussion in the community.
In particular, recently appeared a critical reply where pointed some thermodynamical inconsistencies of the derivations in our paper (J. Chem. Theory Comput., 11, 378–380).
In the present work we re-derive the pressure correction in a more simple way and show that despite the inaccuracies during the derivation, the final form of the previously derived correction is correct. 
We also test the applicability of the proposed correction to the functionals which include a three- and many- body terms from the fundamental measure theory (FMT) for hard sphere fluid.
We test all the functionals on a set of model systems and discuss the obtained results.

\end{abstract}

\section{Introduction}

\subtopic{liquid state theories (see blabla from Volodymyr and from Daniel)}

Solvation Free energy fundamental quantity in physical chemistry which allows ones to compute many useful properties 
of substances in solution
Unfortunately, the solvation free energy calculations are computationally expensive.
That's why there exist alternative, faster but sometimes less accurate methods, like continuum electrostatics models\cite{warwicker_calculation_1982} or energy representation technique\cite{matubayasi_theory_2000}. 
In this group of methods, the methods based on the Classical Density Functional Theory (DFT) are one of the most promising techniques for solvation free energy calculations. They are orders of magnitude faster than the simulations bu nevertheless preserve an important information about the solvent structure around the solute\cite{hansen_theory_2000}.

\subtopic{HNC $=>$ bad SFE $<=$ high pressure (ljguy, Evans, Guillaume).}

However, the accurate solvation free energy calculation in the Classical DFT approach appeared to be a challenging task.
Although the methods based on the simplest Hyper Netted Chain (HNC) approximation \cite{morita_theory_1958,hansen_theory_2000} can give a qualitatively correct estimation of the solvent structure around the solute the calculated values of the solvation free energies are unrealistic and far from experimental ones \cite{lue_liquid-state_1992}. 
It was noticed, that the error in the solvation free energy calculations is proportional to the partial molar volume of the solute\cite{chuev_improved_2007}. 
There were proposed the empirical correction models for the RISM and 3DRISM, which used this observation and were able to predict the solvation free energies of simple solute with accuracy of 1 kcal/mol\cite{palmer_toward_2011,ratkova_accurate_2010,palmer_accurate_2010}.  
In our recent paper we were able to show that these empirical corrections can be explained by the difference in pressure in HNC approximation and in experiments\cite{sergiievskyi_fast_2014}.

\subtopic{ bad SFE $=>$ HSB correction (Wu) $=>$ yes but what is the pressure then? Should it be corrected? If yes then SFE's not correct anymore.}

Another approach to the solvation free energy calculations is to improve the HNC model itself.
The aim of improvements is to include the three and many body interactions to the model.
Some of the models include the empirical terms for improvement the water structure \cite{du_solvation_2000,jeanmairet_molecular_2015,jeanmairet_molecular_2013}, another use the three and many-body interactions from the simplified models .
The models which use the experimental c-functions, and include the many-body interactions from the Fundamental Messure Theory (FMT) are called FMSA or the models with the Hard Sphere bridge \cite{rosenfeld_free-energy_1989,kierlik_free-energy_1990}.
It was shown, that using these models it is possible to predict accurately the solvation energies of alknes \cite{levesque_scalar_2012}, and also more complex solutes \cite{liu_high-throughput_2013}.
One of the disadvantages of these models is the fact that they still contain an  empirical parameter:  hard sphere diameter of the solvent.
The result of the calculations are very sensitive to this parameter and the optimal choice of the proper solvent diameter is a challenging task. 
Another disadvantage of this kind of empirical methods is that the predictability of the model with fitting parameters is always a question, and need to be proved by the tests on extensive number of systems, which is computationally demanding and never give a full understanding of the  model limitations. 
In a light of our previous paper it would be also interesting to measure the pressure in these models, to answer the question whether it is realistic or does the HS diameter is just another fitting parameter. 

\subtopic{ high P $=>$ Volodymyr's correction }

Although the explanation given in our previous paper is not perfect and we agree with the derivations given in \cite{chong_thermodynamic-ensemble_2015} we still cannot agree with the conclusion of that paper that the proposed correction does not have a physical grounds.
We show below that the final formula for the solvation free energy calculation  given in \cite{sergiievskyi_fast_2014} can be correctly explained by a simple difference in pressure in the HNC model (which is extremely high) and atmospheric pressure which is almost negligible in comparison to it. 
So, the numerical results of the paper \cite{sergiievskyi_fast_2014} are still correct and transferable to any Classical DFT calculations with HNC functional.

\subtopic{ Can one use HSB to correct both the pressure and the SFE?}

Additionally, we derive the similar pressure expression for the functionals with the HS bridge and demonstrate that the sensitivity of the model to the hard sphere radius is also due to the pressure of the model. 
We discuss the different ways to choose the solvent hard sphere diameter which can either reproduce the correct pressure or the correct free energies of the small solutes (with the price of loosing the physical consistency). 
At the very end we try to find the systematic components in the expressions which reproduce the correct pressure, and show that for huge variety of systems it can be explained with a  good accuracy by adding the ideal-pressure correction, which poses new challenges for the theoretical justification of this correction.

\section{ Pressure correction in classical DFT and related methods}

\begin{figure}
\center
\includegraphics[width=0.5\textwidth]{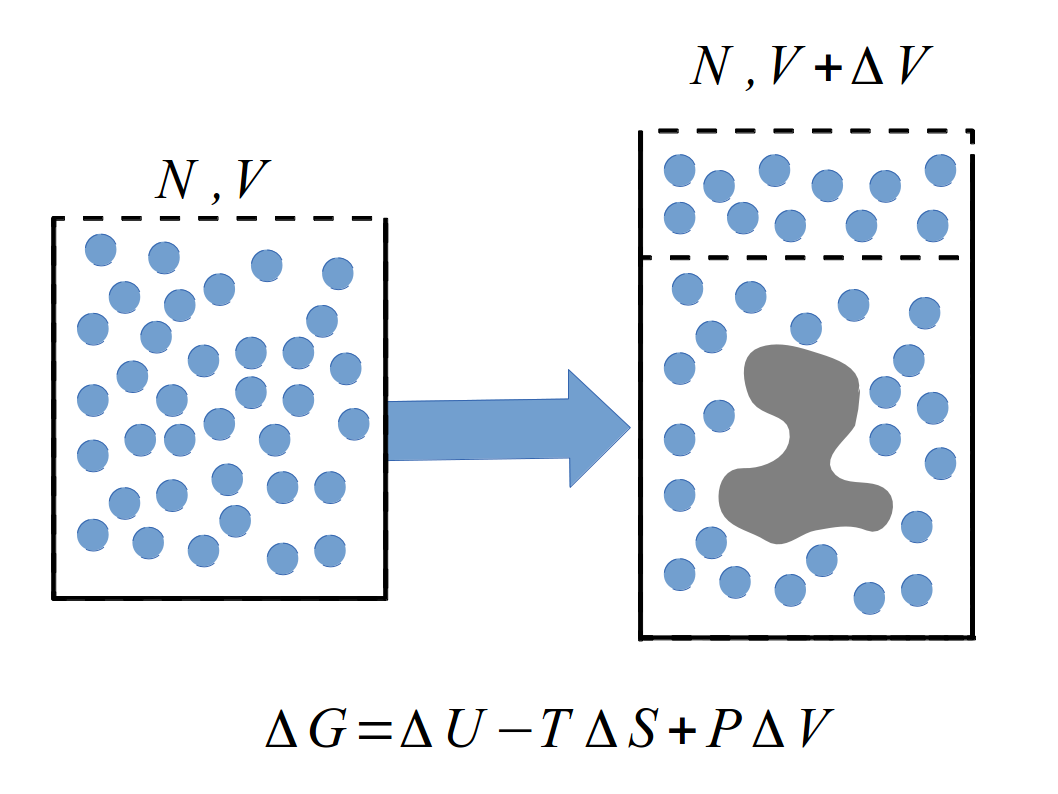}
\caption{ 
\label{fig:DeltaG}
Solvation process in the NPT ensemble and three contributions to the Gibbs energy change.
}
\end{figure}

\begin{figure}
\includegraphics[width=1\textwidth]{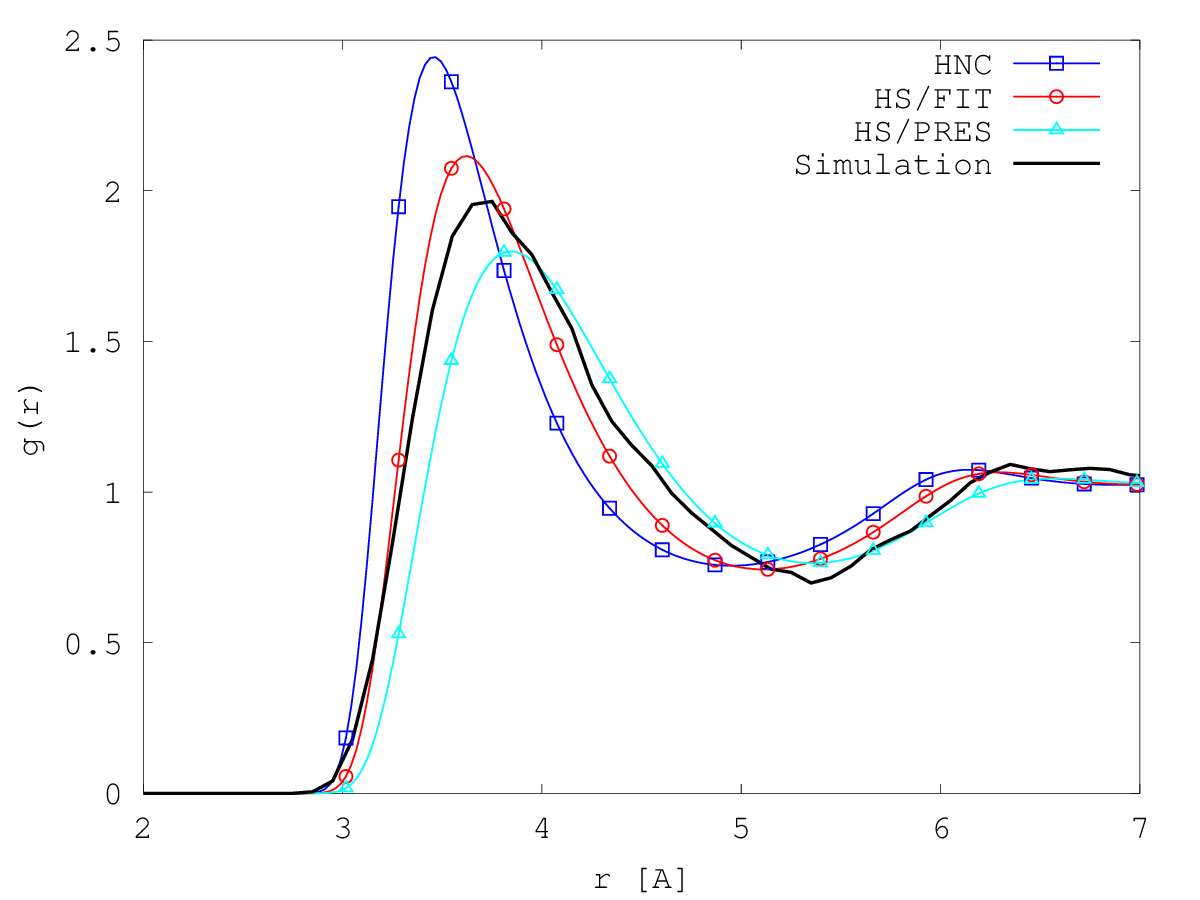}
\caption{\label{fig:g_methane}  
Radial distribution functions of water around methane computed with different Hard Sphere radius parameter. 
}
\end{figure}

Let us consider the process of solvation of the rigid solute.
As it was absolutely correctly mentioned in Ref. \cite{chong_thermodynamic-ensemble_2015}, it does not matter which kind of process (NVT, NPT, $\mu VT$) do we consider, we will always recover with the same solvation free energy.
So, let us consider the solvation process in $NPT$ ensemble, where the pressure, temperature and the number of particles are constant as this ensemble is  commonly used in experiments and simulations.
By definition in the solvation process the solute is transfered from the fixed position in the ideal gas to the fixed position in the solvent \cite{ben-naim_solvation_1984}.
Then during the solvation process the volume of the liquid system changes by some volume $\Delta V$, which is (by definition) the partial molar volume of the solvent (see Figure \ref{fig:DeltaG}). 
According to the basic thermodynamic relations we can find the components of the Gibbs solvation energy $\Delta G$:
\begin{equation}
\label{eq:DeltaG}
   \Delta G = \Delta F + P\Delta V  
\end{equation}
where 
$\Delta F = \Delta U - T \Delta S$ is the Helmholtz free energy change, P is the pressure. 
Different properties of the liquid contribute to the $\Delta F$ and $P\Delta V$ components of the solvation free energy.  
Both: internal energy change $\Delta U$ and the entropy change $\Delta S$ depend in the first turn on the structural changes in the solvent, but not directly depend on the volume change $\Delta V$. 
In turn, the pressure term $P \Delta V$ is responsible for the energy change due to the volume change, while it does not depend on the structure of the solvent.
Separation of  the energy contributions in this way allows us to propose the reasonable approximations for the solvation free energy $\Delta G$.

It is known, that the methods based on the Classical DFT approach can qualitatively correctly predict the solvent structure around the solute 
\cite{
beglov_numerical-solution_1995,
beglov_integral_1997,
hirata_molecular_2003,
levesque_solvation_2012,jeanmairet_molecular_2013-1}.
For example, in Figure \ref{fig:g_methane} are shown the radial distribution functions of water around the methane in water.
Although the quantitative behavior is not so good, and there is a series of well-known problems of the Classical DFT methods, we may still assume that the the structure predicted by the Classical DFT methods is good enough to roughly reproduce the structure-based component $\Delta F$ of the Solvation free energy.

In the other hand, it is also known and shown in our previous paper \cite{sergiievskyi_fast_2014}, that the models based on the Homogeneous Reference Fluid approximation, like HNC models, fail to predict the correct pressure. 
As we show below, the typical pressure in HNC approximation with SPCE solvent at room temperature and usual water density is about 11.5 KBar, which is 4 order of magnitude higher than in experiments. 
In this situation it is reasonable to expect that the pressure contribution is a dominant contribution to the solvation free energy and the main source of the errors of the model. 

To correct these errors we can have several approaches.
The simplest one is an \emph{a posteriori pressure correction}.
Writing equation \eqref{eq:DeltaG} for both: DFT calculations and experiments and using the approximation $\Delta F_{DFT} \approx \Delta F_{expt}$ we get the approximation for the experimental solvation free energy $\Delta G_{expt}$:
\begin{equation}
\label{eq:PC}
  \Delta G_{expt} \approx \Delta G_{DFT} - P_{DFT} \Delta V_{DFT} 
   + P_{expt} \Delta V_{expt}
\end{equation}

Another approach is to use the advanced free-energy functional.
As the functionals are rarely completely parameter-free, we have to possibilities: either to fit the parameters to get the correct SFE of some given molecule (set of molecules), or take the parameters which reproduce the correct pressure. 
(The ideal functional would do both.) 

In the next sections we consider the described approaches and discuss the numerical results.

\subsection{HNC and a posteriori Pressure Corrected HNC}


\begin{figure}[h!]
\includegraphics[width=1\textwidth]{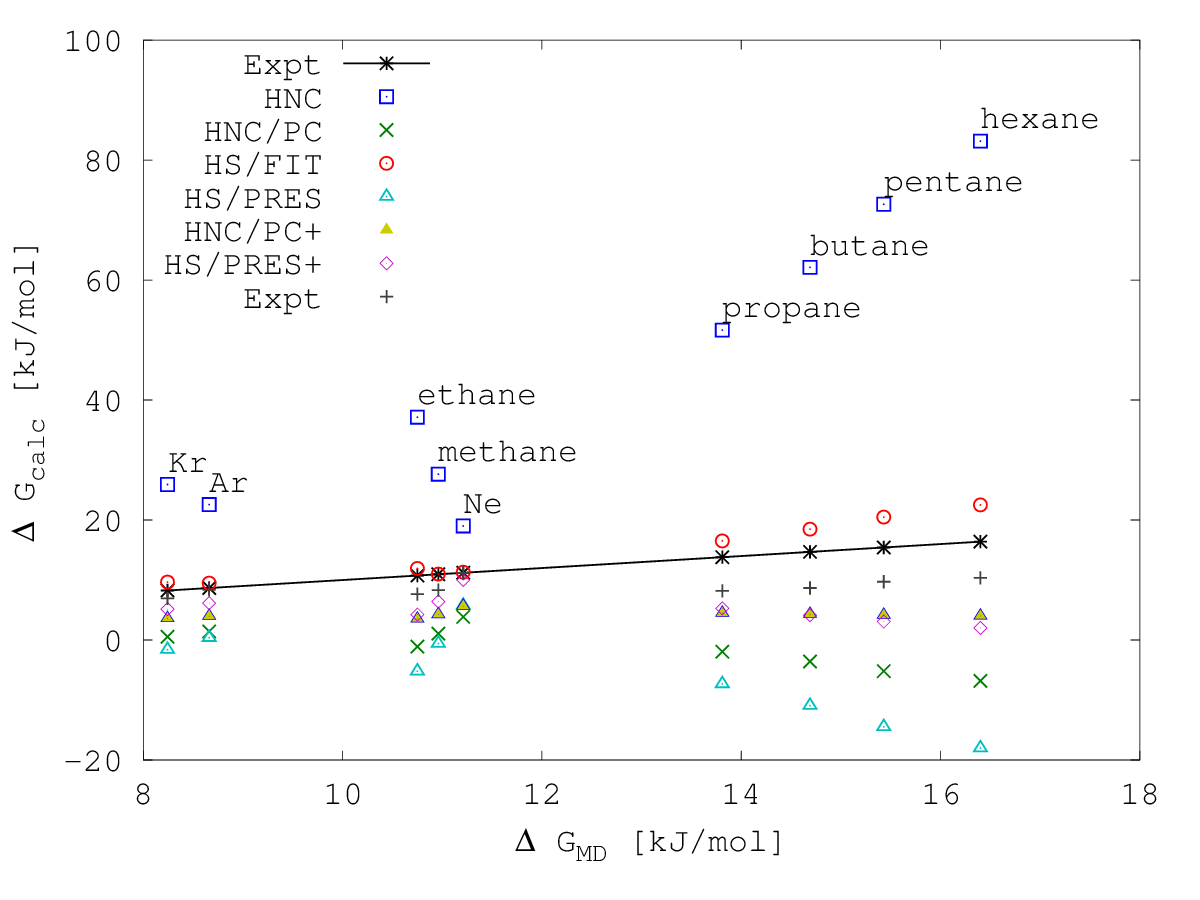}
\caption{\label{fig:SPCE} Results of the MDFT calculations for the small Alkanes and rare gases. HNC corresponds to the Homogeneous Reference Fluid approximation, HNC/PC is the same but includes the pressure correction, HS/FIT corresponds to the calculations with  the Hard Sphere bridge, where the Hard Sphere radius is chosen to reproduce the SFE of methane, HS/PRES is the same, but the Hard Sphere radius is chosen to reproduce the Atmospheric pressure, HNC/PC+ and HS/PRES+ are the results with additional $\rho kT \Delta V$ term. }
\end{figure}

One of the simplest and most popular models considered in the  classical DFT theories is the Hypper Netted Chain (HNC) or Homogeneous Reference Fluid (HRF) approximation.
In this model the excess free energy functional $\mathcal{F}^{exc}$ is expressed as a second-order Taylor series approximation around the homogeneous fluid density $\rho_0$.
Subtraction of the $\mu N$ term with $\mu = \delta \mathcal{F}^{exc} / \delta \rho$ and reference fluid free energy leads to the following Grand potential functional with respect to the reference fluid:
\[
 \Delta \Omega[\rho] = 
 \Omega[\rho] - \Omega[\rho_0]
 =
  kT
 \int
  \left[
  \rho(1) 
  \ln { \rho(1) \over \rho_0 }
  - \Delta \rho(1)
  \right]
  d1
  + 
  \int U(1) \rho(1) d1
  -
  {kT \over 2}
  \iint  \Delta \rho(1) c(12) \Delta \rho(2) d1d2
\]
where arguments $1$, $2$ stay for the positions and orientations of the solvent molecules,
$U$ is the potential of the solute molecule,
$\Delta \rho(1) = \rho(1) - \rho_0$,
$c(12) = -\beta \delta^2 \mathcal{F}^{exc} / \delta \rho(1) \delta \rho(2)$ is a pair direct correlation function,
$k$ is a Boltzmann constant, $T$ is a temperature, $\beta=1/kT$.
By minimizing the functional over the solvent density $\rho(1)$ one can find both: the solvation free energy $\Delta \Omega$ and the density distribution $\rho(1)$.

\subtopic{ figure : SFE of small hydrophobic molecules (rare gaz and alkanes) : MD vs HNC. $=>$ bad. That's known.}

It is known that without the pressure correction HNC calculations dramatically overestimate the solvation free energies of the solutes \cite{lue_liquid-state_1992,chuev_improved_2007}.
To demostrate that fact we performed the HNC calculations for small hydrophobic solutes in water (see Figure \ref{fig:SPCE}, squares).
As expected, the HNC SFE predictions are always too high, sometimes by 200-400\%.

\subtopic{ figure : SFE/(4 pi R$^2$) as a function of the radius R of a growing something (HS or LJ) }

\begin{figure}[h!]
\includegraphics[width=1\textwidth]{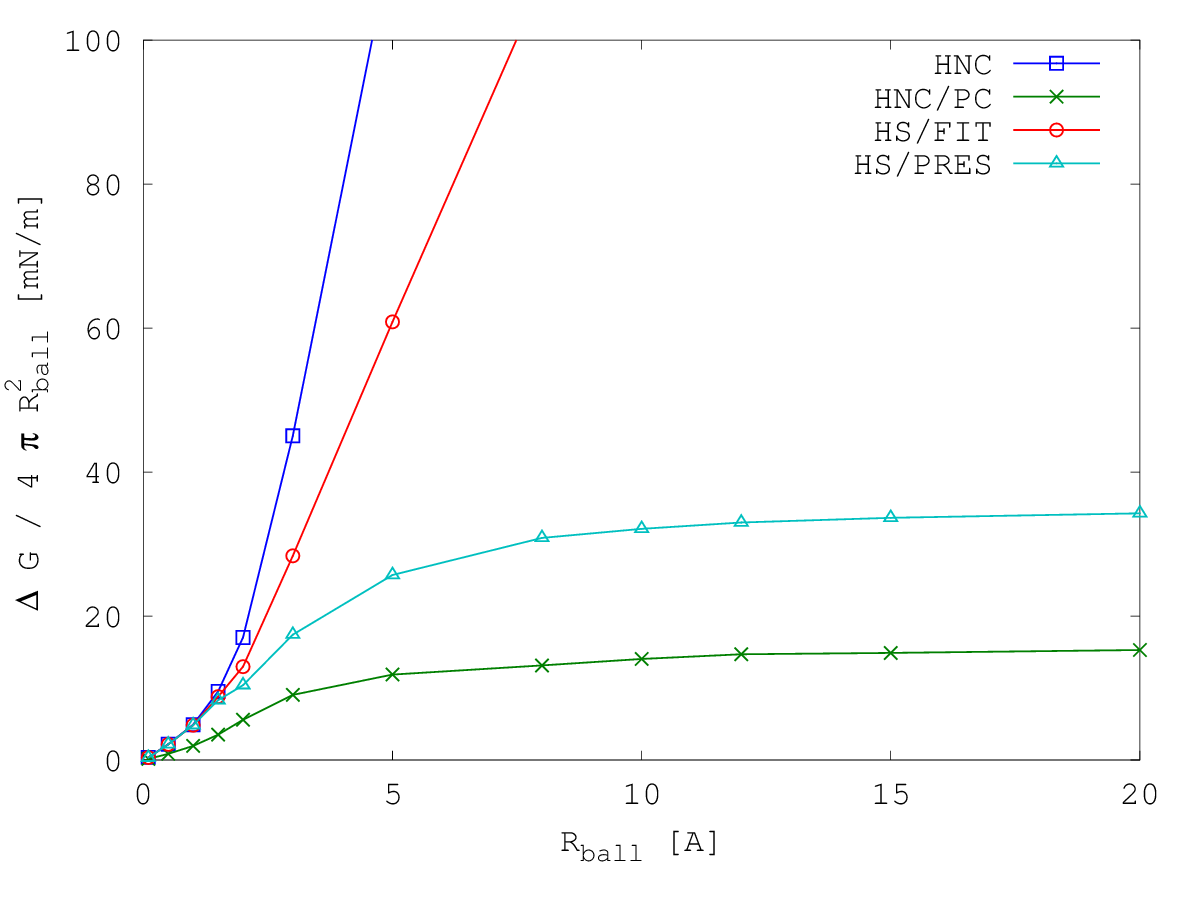}
\caption{\label{fig:SurfTens} Dependence of the free energy per surface area for the Hard Spheres of different diameter immerced in water. Results are shown for different free energy functionals. The abbreviations are the same as in Figure \ref{fig:SPCE}  }
\end{figure}

To check whether the errors are random or systematic, we performed the calculations for the simpler system: hard sphere solutes in water.
It is known, that for the macroscopic solutes the free energy is proportional to the surface of the solute.
In case of the liquid-vapour interface the proportionality coefficient is the surface tension $\gamma$.
In Figure \ref{fig:SurfTens} we see the dependency of the Solvation Free Energy per surface area on the hard sphere diameter. 
We see that HNC functional fails to produce the correct behavior of the surface tension: the solvation free energy per surface area in HNC approximation is proportional to the hard sphere radius and never comes to the plateau.
So, HNC fails to reproduce even the qualitative behavior of the solvation free energy for the simple model system.

To reduce the errors, we can use the a posteriori pressure correction described in the previous section.
The pressure in the HNC approximation can be found using the known thermodynamic relation for the homogeneous fluid:
\[
  \Omega[\rho_0] = - P V
\]
If we set zero density to the functional, we get:
\begin{equation}
\label{eq:PHNC}
P_{HNC}  = \Delta \Omega[0] / V = -\Omega[\rho_0] / V = \rho_0 kT -{kT \over 2} \rho_0^2 \hat c(k=0)
\end{equation}
where
$ 
 \hat c(k=0) = \int c(12) d1
$.
The definition of the pressure, given above corresponds to the results in Refs. \cite{rickayzen_integral_1984,powles_density_1988}.
We should say, that this is notthe only possible definition.
To avoid misunderstanding in the community, we briefly discuss the different ways of pressure definition in Classical DFT and their domain of applicability.

The pressure computed in \eqref{eq:PHNC} is a so-called \emph{compressibility route pressure}, which differs from the \emph{virial route pressure}.
We emphasize, that it is consistent in our derivations to use \emph{compressibility}, but not \emph{virial} route pressure. 
To support that, we can write the virial-route pressure in the simplest case of spherical particles \cite{kalikmanov_statistical_2001}:
\[
 P = \rho_0 kT - 
 {2 \pi \over 3} \rho^2 
 \int_0^{\infty} 
  {\partial u(r)  \over \partial r} g(r) dr
\]
where $u(r)$ is a pair-wise potential and $g(r)$ is a radial distribution function of the solvent. 
It is well seen that this quantity depends on the structural properties of the system only, and is \emph{the same} for any choice of the approximate functional $\Delta \Omega[\rho]$. Thus, it cannot be used to correct its deficiencies.
Also, the pressure in \eqref{eq:PHNC} should not be mixed with the pressure which can be computed from solving the HNC equations for solvent, as it is done in Ref. \cite{morita_theory_1960}. In our model the solvent direct correlation function $c(r)$ is extracted from the simulations, so it is inconsistent to use it in any way in the HNC equations for solvent.
In turn, the pressure defined as \eqref{eq:PHNC} correctly defines the energetic behavior of the system, in a sense that it shows the energy change then the volume of the system changes, and this is exactly the quantity we are interesting in. 

For the SPCE water model used in our simulations the compressibility route pressure computed by \eqref{eq:PHNC} we get the the pressure of 11.477 KBar. As this pressure is 4 orders of magnitude higher than ones used in experiments, we can neglect the $P_{expt} \Delta V_{expt}$ term in \eqref{eq:PC}.
In that case the solvation free energy in the HNC approximation with the pressure  correction (HNC/PC) can be written as
\[
   \Delta G_{HNC/PC} = 
   \Delta \Omega[\rho] 
   - \rho_0 kT( 1 - {\rho_0 \over 2} \hat c(k=0) \Delta V_{HNC}
\]
 This result is completely consistent with the result which we got in Ref \cite{sergiievskyi_fast_2014}, (eq (11)), which unfortunately had not the best reasoning.

Looking at Figure \ref{fig:SPCE} we see, that the results for HNC/PC improves the solvation free energies of small solutes (rare gases and methane), but it essentially underestimates the solvation energies of larger Alkanes. 
Also, as it can be seen in Figure \ref{fig:SurfTens}, the HNC/PC model is able to reproduce the qualitative behavior of the free energy of a big hard sphere solute in water. 

\begin{figure}
\includegraphics[width=1\textwidth]{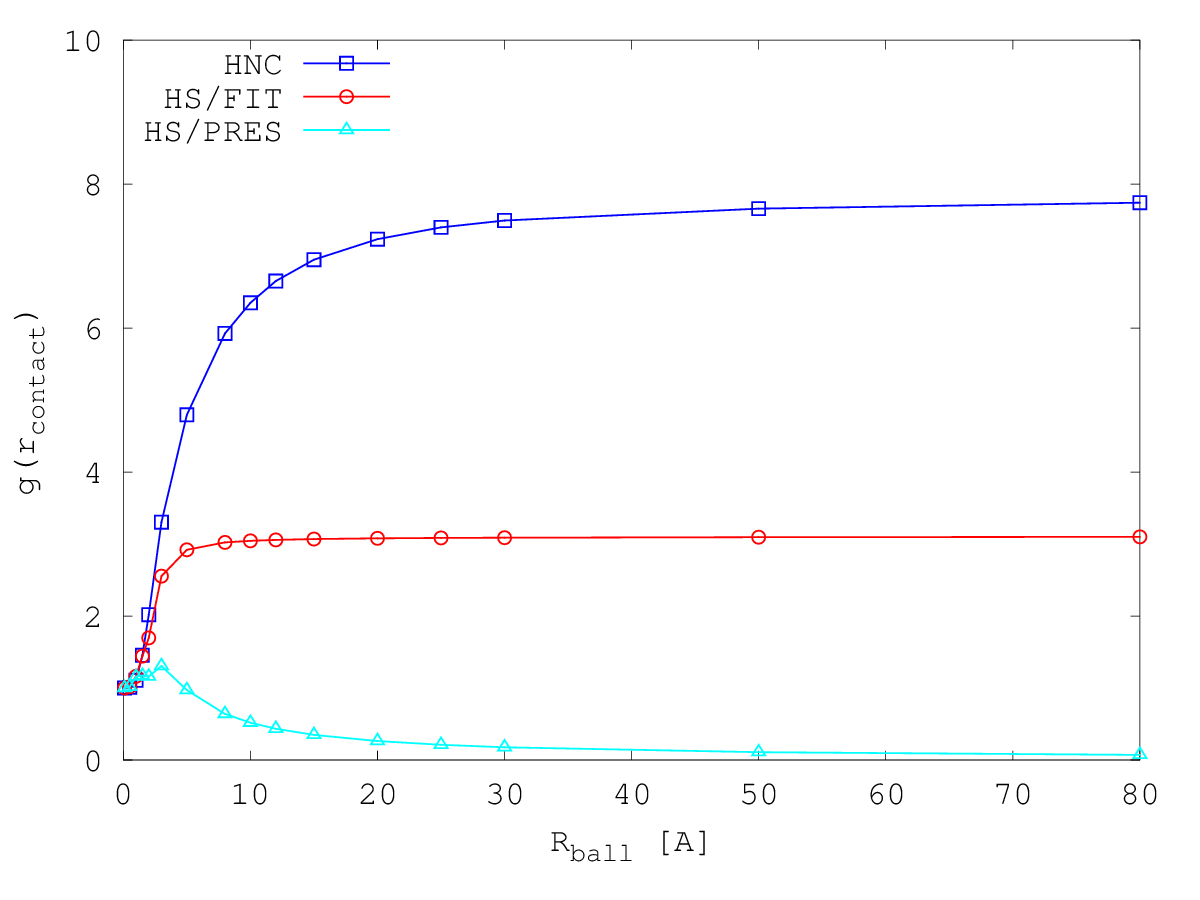}
\caption{\label{fig:g_contact} 
Contact density and the maximum density as a function of hard sphere radius for different free energy functionals. 
Note: the contact density coincides with the maximum density for HNC and HS/FIT functionals.
The abbreviations are the same as in Figure \ref{fig:SPCE}
}
\end{figure}

\begin{figure}[h!]
\includegraphics[width=1\textwidth]{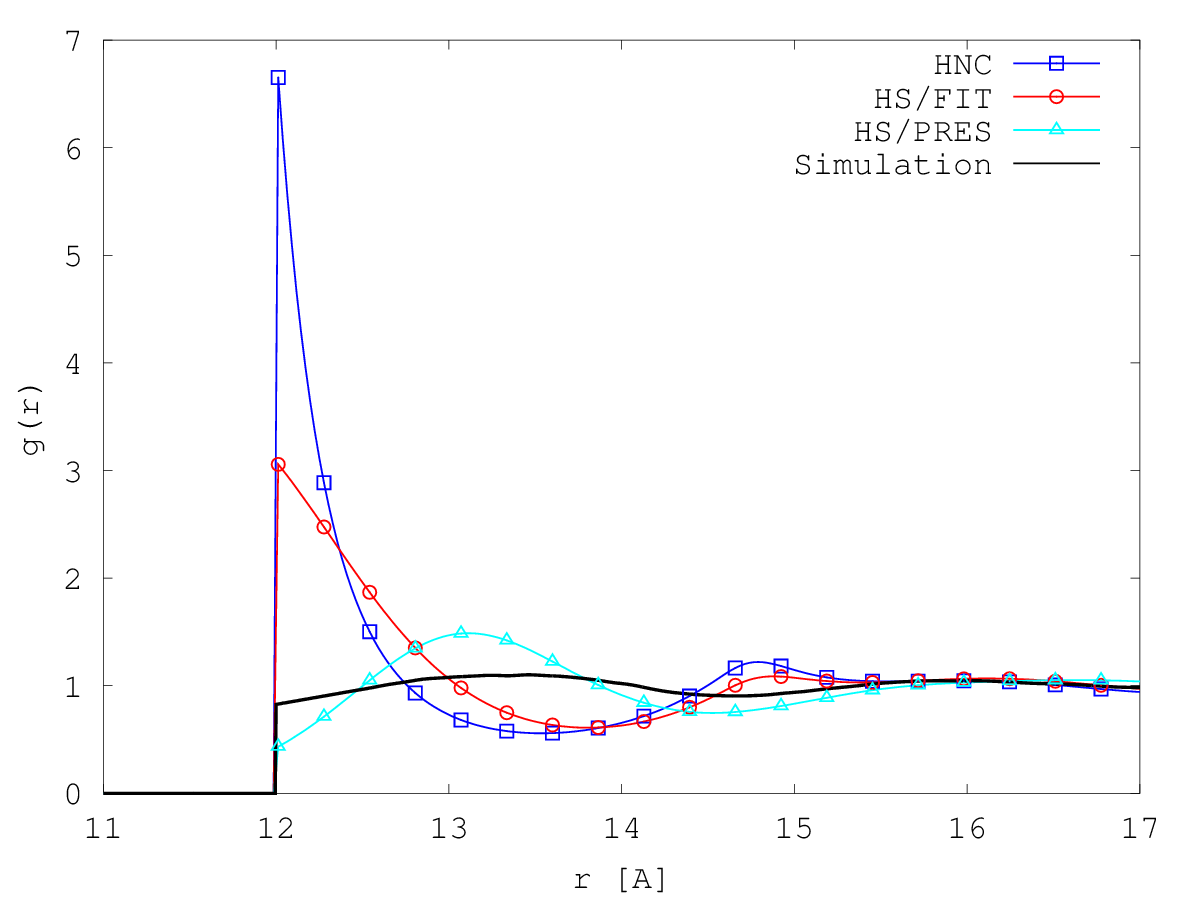}
\caption{ \label{fig:gr_R12} 
Density distribution functions of water around the Hard Sphere of radius R=12 \Angstr, calculated with three different functionals.
The functional name abbreviations are the same as in Figure \ref{fig:SPCE}
 }
\end{figure}

\subtopic{ figure : g(r,R) $=>$ or 10 g(r) \maxcomment{I will produce myself} }
\voovcomment{Don't you think that the Figure \ref{fig:gr_R12} is more representative? the behavior is actually similar for any size of the ball, and in Figure \ref{fig:gr_R12} we see all the methods together, can comare them etc.
And one can look at Figure \ref{fig:g_contact} to see what is the contact density for each size.
}

Despite the fact that pressure correction can to some extent correct the misleadingly high error of the HNC model, it is still far from being perfect.
The assumption used in the pressure correction approximation is that HNC can give a relatively good approximation of the solvent structure around the solute. Although in many cases this assumption is quite reasonable, in some situations the structure in HNC approximation is very far from reality. 
One of such cases is the contact density of the water with a hard-sphere.
According to the 
so-called contact value theorem the contact density near the hard wall is proportional to the pressure \cite{warwicker_calculation_1982}.
 Note, that for the compressibility route pressure and approximate functional it should not necessarily hold, but, a high contact density can be another indication of a functional failure to reproduce the simulation and experimental results.
 
\subtopic{ ?figure : contact density as a function of the radius R }

In Figure \ref{fig:g_contact} the contact density for the different hard sphere diameters is presented.
We see that the HNC density grows up to $g(r_{contact}) = 7.7$. The real contact density should decrease with the hard sphere diameter increase, and tend to the small value which corresponds to the atmospheric pressure  $P/kT \approx 0.01 $.
The water RDF for the $R_{ball}$ = 12 \Angstr is shown also in Figure  \ref{fig:gr_R12}. We see that the radial distribution function (RDF) calculated with the HNC functional shows completely wrong behavior near the hard sphere surface. So, we may expect also a big errors in calculation the thermodynamic quantities as well. 
So, to conclude: it is known and proven many times that HNC functional is not perfect. 
It predicts the solvation free energies proportional to the volume, and thus fails to reproduce the surface tension. 
Although the pressure corection can correct the unnaturally high values and make them more reasonable, still the deficiencies of the HNC functional do not allow to improve the method further.
HNC fails to reproduce the correct RDF, and in many cases the difference is drastic, so we cannot expect the correct thermodynamics.



%
%
%
%
%
%



\subsection{ Hard Sphere Bridge functional with empirical solvent diameter }


As it was discussed above, there is an alternative way to improve the solvation free energy calculations. To do it one can use an improved functional, which includes many-body terms. 
In our paper we use the mixed functional approach: use the two-body $c$-function extracted from the simulations but add the many-body terms from the model where the analytical solution is known.
The good candidate is the fundamental measure theory (FMT) with the functional for the hard sphere fluid. 
In that case we define the full free energy functional in the following way:
\begin{equation}
\label{eq:FHS}
  \Delta \Omega_{HS}[\rho]
  =
  \Delta \Omega [\rho] 
 + 
  \mathcal{F}_{HS}^{(3+)}[\rho]
\end{equation}
where 
$
 \mathcal{F}_{HS}^{(3+)}
$ 
is a many body interactions of hard sphere fluid, which includes terms starting from the 3$^{\rm rd}$ order and higher:
\begin{equation}
 \mathcal{F}_{HS}^{(3+)}[\rho]
 =
 \mathcal{F}_{HS}^{exc}[\rho]
 -
 \mathcal{F}_{HS}^{exc}[\rho_0]
 -
 \left.
 { \delta \mathcal{F}_{HS}^{exc}
   \over
   \delta \rho
 }
 \right|_{\rho_0}
 \int \Delta \rho(\mathbf{r}) d\mathbf{r}
 +
 {kT \over 2}
 \iint
 \Delta \rho(\mathbf{r}_1)
 c_{HS}(|\mathbf{r}_2 - \mathbf{r}_1|)
 \Delta \rho(\mathbf{r}_2)
 d\mathbf{r}_1 
 d\mathbf{r}_2
\end{equation}
\begin{equation} 
c_{HS}(|\mathbf{r}_2  - \mathbf{r}_1| )
=
-\beta 
\left.
{  \delta^2 \mathcal{F}^{exc}_{HS} 
  \over
   \delta \rho(\mathbf{r}_1) 
   \delta \rho(\mathbf{r}_2)
}
\right|_{\rho_0}
\end{equation}
In the paper we use the Percus Yevick functional \cite{rosenfeld_free-energy_1989} in the scalar representation \cite{kierlik_free-energy_1990}:
\begin{equation}
  \mathcal{F}_{HS}^{exc} 
  =
  \int \Phi(\mathbf{r}) d\mathbf{r}
\end{equation}
\begin{equation}
 \Phi(\mathbf{r}) = 
 -n_0 \ln (1-n_3) 
 + { n_1 n_2 \over 1-n_3}
 + { 1 \over 24 \pi } 
   { n_2^3 \over (1-n_3)^2}
\end{equation}
where $n_i(\mathbf{r}) = \int_V \rho(\mathbf{r}') \omega_i(\mathbf{r} - \mathbf{r}') d\mathbf{r}'$
and spherically symmetric weight functions $w(r)$ are defined in the $k$-space as follows:
\begin{equation}
 \hat \omega_0(k) = \cos kR + {kR \over 2} \sin kR
\end{equation}
\begin{equation}
 \hat \omega_1(k) = {1 \over 2k} ( \sin kR + \cos kR )
\end{equation}
\begin{equation}
  \hat \omega_2(k) = {4 \pi R \over k} \sin kR
\end{equation}
\begin{equation}
  \hat \omega_3(k) = {4\pi \over k^3} ( \sin kR - kR \cos kR )
\end{equation}
where $R$ is the radius of the hard sphere.

\begin{figure}
\includegraphics[width=1\textwidth]{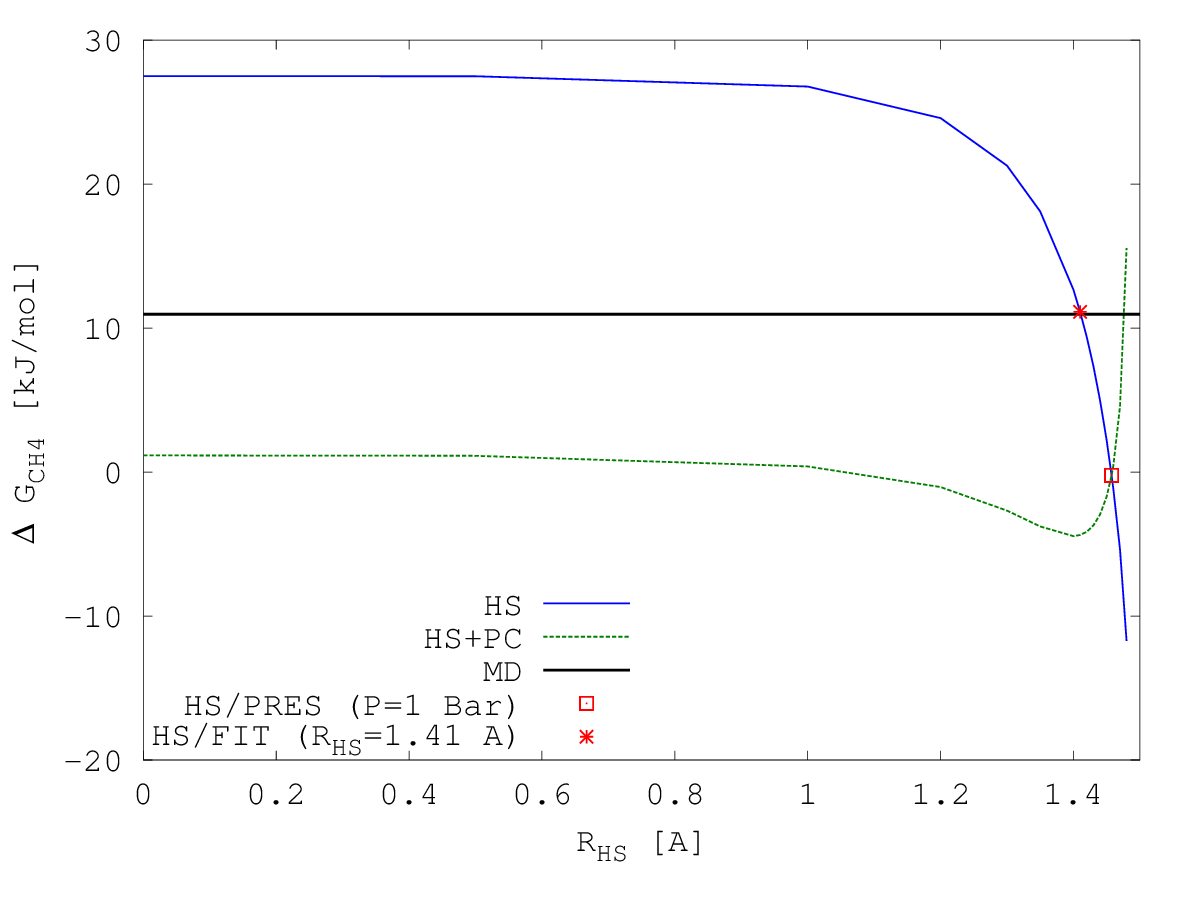}
\caption{\label{fig:SFE_CH4} Free energy of methane calculated with Hard Sphere bridge functional with different hard sphere radii (HS).
HS + PC - the same values with the Pressure Correction, where the  pressure contribution to the free energy is subtracted. 
  }
\end{figure}

The hard sphere radius is not pre-defined for the aqueous solution, so we can consider it to be a free  parameter.
In Figure \ref{fig:SFE_CH4} the dependence of the Solvation Free energy of methane on $R$ is shown.
We see that free energy of solvation depends on the hard-sphere radius $R$, especially in the region $1.2 \Angstr < R < 1.5 \Angstr$ where solvation free energy is extremely sensitive to it and change by 40 kJ/mol.
So, we can choose $R$ which reproduces the solvation free energy of methane. 
In our case it is $R\approx 1.41 \Angstr$.
If the fix this parameter, we obtain the functional which can be used to calculate solvation energies of arbitrary molecules. We refer this functional as HS/FIT.

\subtopic{ same figures as section 2 }


Looking at Figure \ref{fig:SPCE} we see, that HS/FIT functional is able to reproduce the MD solvation free energies with a good accuracy.
It is almost perfect for the small one-atom molecules. For larger Alkanes there is a growing deviation, which is still much smaller than HNC/PC results. 
In principle, as it was demonstrated by Wu \cite{liu_high-throughput_2013}, using this method with a proper parameterization it is possible to reproduce the SFE for a larger sets of molecules with a quite high accuracy. 
In Figure \ref{fig:g_methane} we see that HS/FIT method reproduces the height of the first peak of methane-water RDF almost correctly.
However, it is too shallow, and the position of the first minimum is incorrect. 
But in general we may conclude that the results for the small solutes are much better than HNC and nearly of the accuracy of simulations.

However, result for the Large Hard Sphere Solute in water are different. 
In Figure \ref{fig:gr_R12} we see that the 
HS/FIT functional does not reproduce the correct solvent structure around the hard sphere solute.
In Figure \ref{fig:g_contact} we see, that the HS/FIT functional gives the contact density smaller than HNC, but still too high to correspond to real pressure.
The fact that the HS/FIT pressure is incorrect is even better seen in Figure \ref{fig:SurfTens}, where we see that HS/FIT the solvation free energy is proportional to the volume of the solute.

\subtopic{ conclusion : if one uses the so-called bridge 
functional with a radius parameterized over the sfe of let's say methane, then in fact the true sfe, once the pressure is corrected, is wrong.it is not the effect of correcting the pressure $=>$ bad SFE!}

We can conclude that using the the hard-sphere bridge 
functional with a radius parameterized over the SFE of small molecule (in our case methane), we can recover the SFEs of other small molecules (noble gases and Alkanes) with the good accuracy. 
However, in this way we are unable to recover the correct pressure, which results in the incorrect Solvation Free energy per surface behavior  and incorrect solvent structure for the big hard sphere solutes.
So, also HS/FIT method can be used to reproduce the SFEs of small solutes, one need to remember about its limitations of applicability.



\subsection{ Hard Sphere bridge functional with correct pressure }

Instead of using the HS/FIT functional, we can also construct the functional which reproduces the correct pressure of the system. 
To do it, we find the compressibility-way pressure produced by the functional \eqref{eq:FHS}. As in case of HNC functional, we use the relation $\Delta \Omega_{HS}[0] = PV$, which gives the following pressure\footnote{See Supporting Information for the derivation}
:
\[
  P = \rho_0 kT - {kT \over 2} \rho_0^2( c(k=0) - c_{HS}(k=0) )
   + P_{FMT}^{exc}
\]
where 
\[
  P_{FMT}^{exc} = \rho_0 kT ( 4 B + 6B^2 + 3 B^3)
\]
where $B = \rho W_3 / ( 1-\rho W_3)$, $W_3 = 4/3 \pi R^3$.

\subtopic{
 Pressure as a function of the HSBradius $=>$ for 1 atm, one wants HSBradius=...}
 
 \begin{figure}
 \center
 \includegraphics[width=0.7\textwidth]{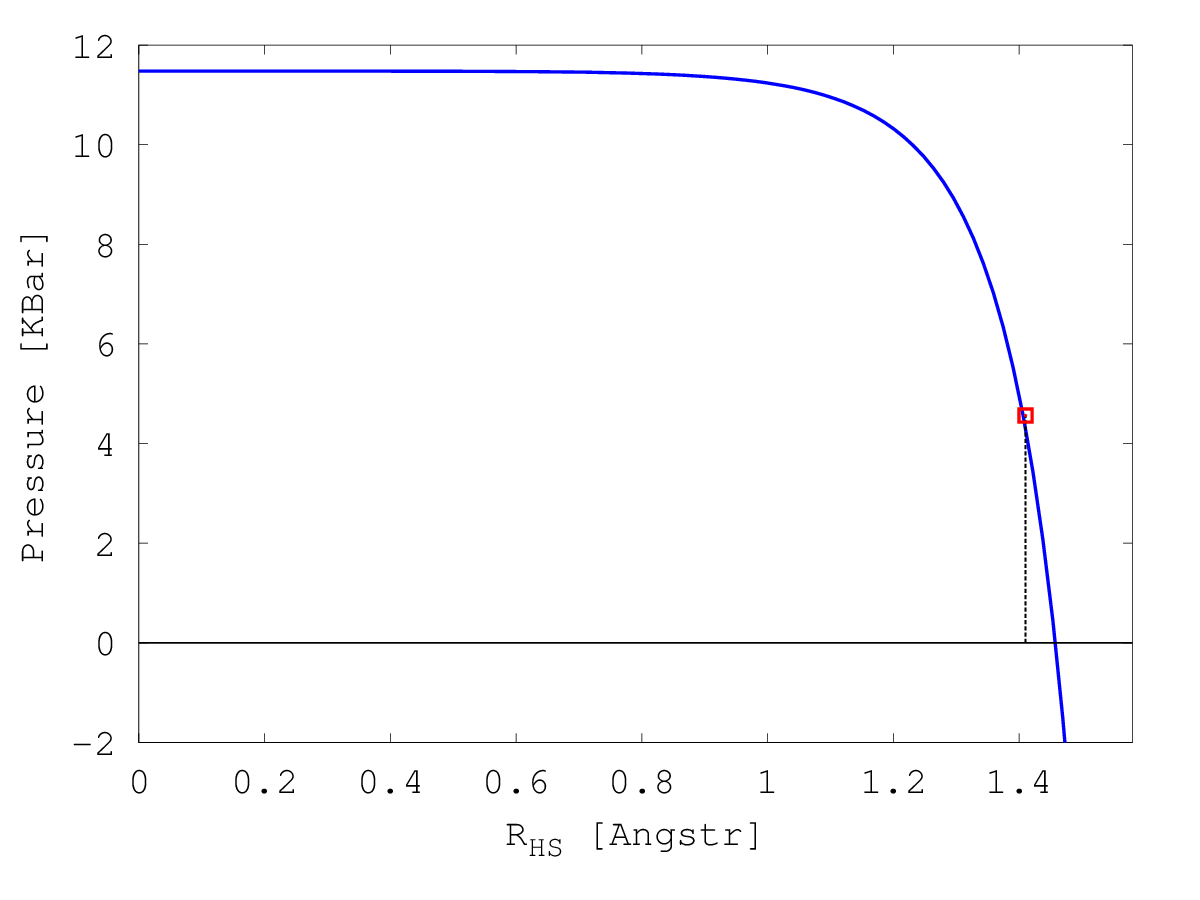}
 \caption{\label{fig:PressureHS} 
 Pressure for the functional with hard sphere bridge as a function of the hard sphere radius of FMT functional.
The pressure of HS/FIT functional with $R=1.41$\Angstr is 4.56 KBar. 
 The pressure of 1 Bar corresponds to $R_{HS}$ = 1.4574\Angstr.
 }
 \end{figure}

This allows us to plot the pressure in HS-bridge functional as a function of $R$.
In Figure \ref{fig:PressureHS} we see that dependency.
We see that for $R<1\Angstr$ the pressure is almost the same as in HNC model, however in region $1.2\Angstr < R < 1.5\Angstr$ the dependency is very strong.
We see that the pressure in HS/FIT model with $R=1.41$ is 4.56 KBar.
The radius $R= 1.4574\Angstr$ corresponds to the atmospheric pressure. We refer the hard sphere bridge functional with with radius as HS/PRES.
 

\subtopic{ same as section 2. $=>$ thermodynamicaly consistant but bad SFE. }

To check whether our derivations are correct and the HS/PRES functional really reproduces the correct pressure of the system, we can look at the Figure \ref{fig:SurfTens}. We see that at large Hard Sphere diameter the HS/PRES free energy is proportional to the surface area, but not to the volume, which is an expected result.
 Now we can check how good can the HS/PRES functional reproduce the solvent structure.
 In Figure \ref{fig:g_methane} we see that the first peak of methane-water RDF is not reproduced completely correct: it is shifted from the methane and is a little bit smaller than in simulations. 
 However, the first minimum and the second peak are reproduced much better than in other methods (HNC, HS/FIT).
 The results for the large hard spheres in water are even better. In Figure \ref{fig:g_contact} we see the correct behavior for the contact density for HS/PRES functional: it tends to a small value when the radius of the hard sphere increases. 
 In Figure \ref{fig:gr_R12} we see, that Hard-sphere water RDF for HS/PRES methods behaves qualitatively correct and much more consistent with the simulations than other methods. 
However, solvation free energy results are not so good. 
In Figure \ref{fig:SPCE} we see that the HS/PRES functional underestimates the free energies of all solutes, it is especially visible for multi-atom Alkanes.

\subtopic{HNC+HSB-that-gives-correct-SFE-of-methane-with-ad-hoc-pressure-correction}

\subtopic{ SFE of methane as a function of HSBradius that takes into account the pressure correction. Sadly, it *always* overshoot the expected correction == SFE too low.} 


As the pressure correction \eqref{eq:PC} can be defined for any functional, we in principle, we can also imagine a hybrid method, which uses both: the pressure correction and the hard sphere bridge. In Figure \ref{fig:SFE_CH4} we see that this functional  underestimates the SFE for any reasonable parameter $R$. 
The only way to recover the correct solvation free energy is to use the $R> 1 \Angstr$, which corresponds to a negative pressure and is thus completely unphysical. 
Unfortunately, with the hard sphere bridge functional we cannot get all: good solvation free energy, correct structure and thermodynamically consistent pressure.
However, we may decide what is more important for us in each concrete case and use the corresponding parameters.

\section{ Additional correction }

\maxcomment{ Secret inside ;) Note: I clearly want to put this correction in the paper since the results get *so good*. We can say we don't have an explanation for it yet, but give several hypothesis if we clearly state they are hypothesis. }

In the previous section after 
analysing the ability of different functionlas to reproduce SFE and testing them on the model system of large hard spheres solvated in water we came to the conclusion that we cannot simultaneously accruratly reproduce all the properties: pressure, SFE and solvent structure, and thus need to look for a compromise. 
However, we need to say that this conclusion is not completely right. 
Doing the numerical experiments for the systems with different parameters of the functionals we noted that the errors in SFE calculation of the functionals which correctly reproduce the pressure (HNC/PC, HS/PRES) is not random.

\begin{figure}
\includegraphics[width=0.8\textwidth]{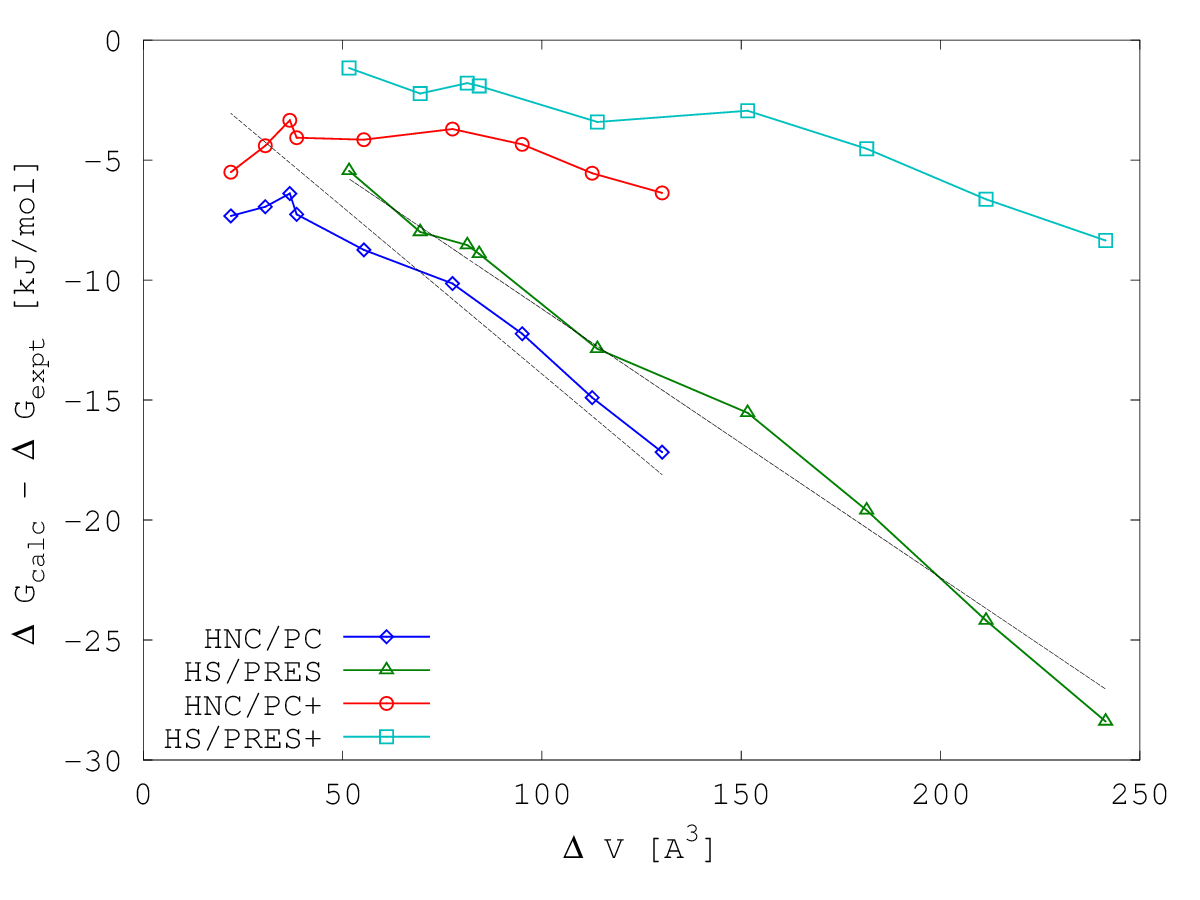}
\caption{
\label{fig:Myst_dV}
Dependencies of the errors of the functionals after the pressure correction on the partial molar volume of the molecules.
HNC/PC+ and HS/PRES+ corresponds to the calculations with the additional $\rho kT \Delta V$ term.
}
\end{figure}

In Figure \ref{fig:Myst_dV} one sees the dependency of this error on the partial molar volume $\Delta V$.
We can spot an evident linear dependency for both of the methods.
The slope of the dependency is almost equal for both of the methods and can be approximated with a good accuracy by the expression $\rho kT$.
Doing the same analysis for the greater number of molecules we can suppose that this is not just a coincidence, but a systematic behavior (See Supporting  Information).
On the basis of our considerations we can propose an additional correction:
\begin{equation}
\label{eq:corr+}
  \Delta G_{PC+} = \Delta \Omega[\rho] - (P - P_{expt})\Delta V + \rho kT\Delta V
\end{equation}
This correction can be used for any functional.
We denote the results where this correction was used by writing an additional ``+'' at the end, for example HNC/PC+, HS/PRES+.
In Figure \ref{fig:Myst_dV} we see that the additionally corrected results are much less dependent on $\Delta V$ and are much more close to the experimental ones.

\begin{figure}[h!]
\includegraphics[width=1\textwidth]{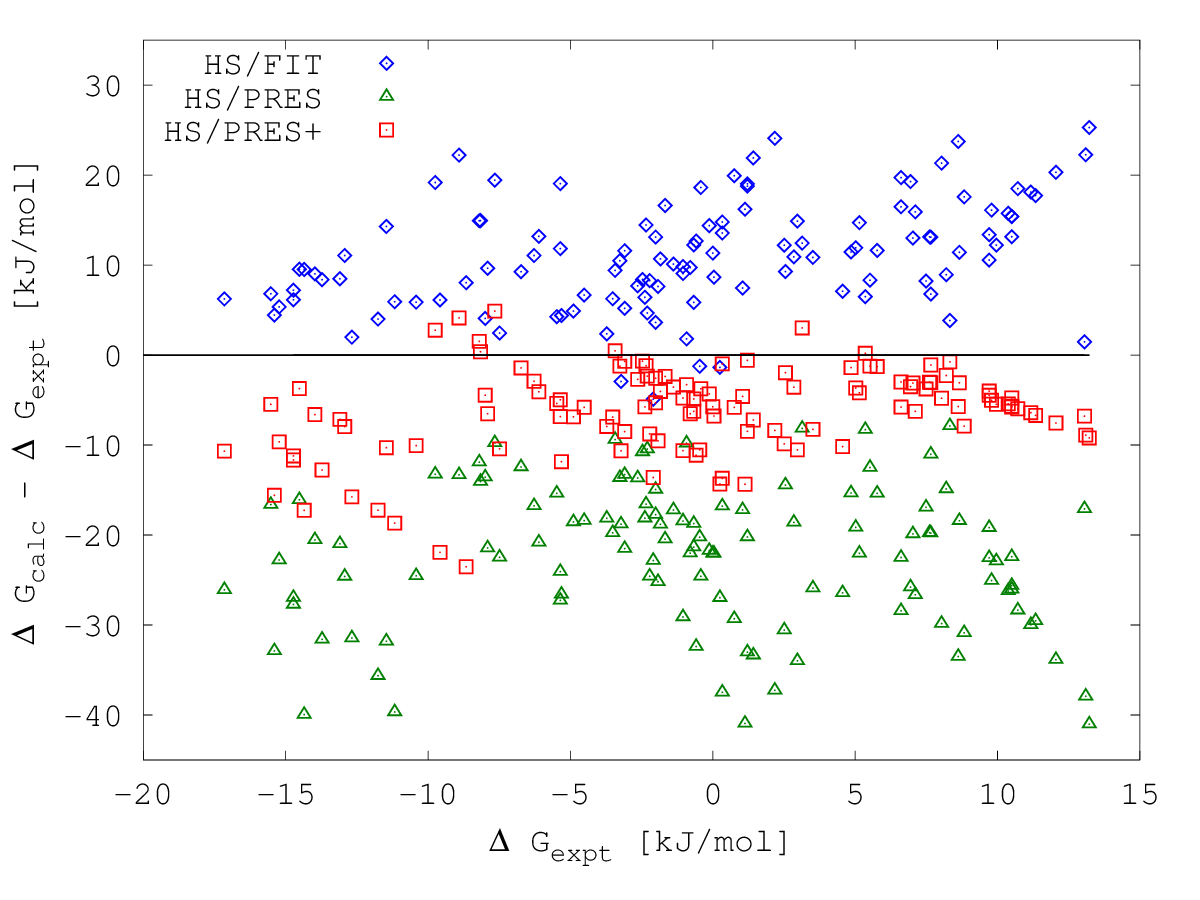}
\caption{\label{fig:rhokT_hs} Error of the free energy calculations for 122 molecules calculated with the free energy functional with the hard sphere bridge. Blue diamonds (HS/FIT) are the results for the $R_{HS}$ = 1.41 \Angstr  which reproduces the free energy of methane.
Green triangles (HS/PRES) are the results for the $R_{HS}$ = 1.4574 \Angstr,
 which reproduce the correct pressure. Red squared (HS/PRES+) are the results which reproduce the correct pressure with additional $\rho kT \Delta V$ contribution added.}
\end{figure}

\begin{figure}[h!]
\includegraphics[width=1\textwidth]{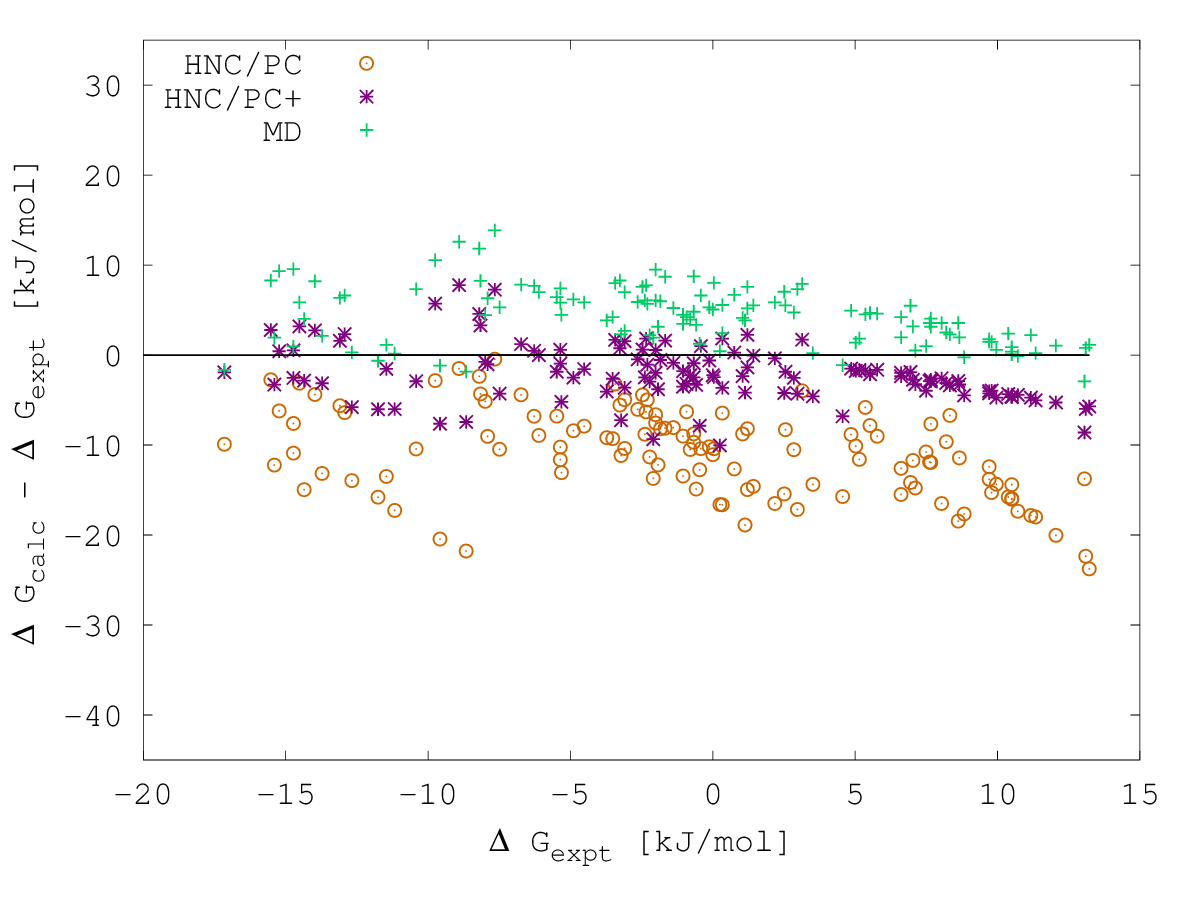}
\caption{\label{fig:rhokT_hnc} 
Error of the free energy calculations for 122 molecules calculated with the Homogeneous Reference Fluid (HNC) free energy functional. Orange circelss (HNC/PC) are the results for Homogeneous Refenrence Fluid approximation with the Pressure Correction.
Violet Stars (HNC/PC + rhokT) are the same result with additional $\rho kT \Delta V$ contribution.
Green pluses (MD) are the results of the MD simulations.
}
\end{figure}

To check the effectiveness of the additional correction we performed the calculations for the larger set of molecules.
In Figure \ref{fig:rhokT_hs}, Figure \ref{fig:rhokT_hnc} we see comparison of the different methods with Pressure Correction for 122 organic molecules
\footnote{Experimental and simulation data taken from \cite{david_l_mobley_small_2009}. See the full list of molecules in Supporting Information. }
.
We see that results with the additional correction can essentially  improve results for both cases: a posteriori pressure corrected HNC functional (HNC/PC+) and HS-Bridge functional with the correct pressure (HS/PRES+).
In the case of HNC/PC+ the resulting error is comparable to the error of simulations.
We see that in general HS/PRES+ gives a more dispersed results than HNC/PC+. 
However, this can be explained by the numerical instability of the HS-bridge calculations for strongly polar molecules.
For non-polar molecules with possitive SFE results of the both functionals are comparable and have a similar accuracy.
So, the results suggest us that the correction \eqref{eq:corr+} is not artificial but possibly have some well physical grounds and can be successfully used in Classical DFT calculations.

We need to comment separately the recent paper Ref \cite{misin_communication:_2015}, where the Pressure Correction method is used with the 3DRISM calculations.
In that paper two variants of the a posteriori pressure correction: with and without additional $\rho kT $ term ( which corresponds to HNC/PC and HNC/PC+ in our definitions).
The numerical results presented in this paper suggests that the HNC/PC produces more reliable results in the 3DRISM case.
As an explanation of this fact it was stated, that 3DRISM is forulated for the Grand Canonical ensemble (in contrast to MDFT, as it could be understood from the context).
Although we do not deny the computational results of Misin \& coathours, we need to note, that the explanation of the difference in MDFT and 3DRISM results, given in the above mentioned paper is not good enough.
In particular, for the case of spherical solvent 3DRISM equations are completely equivalent to the MDFT and can be obtained from the later by taking the functional derivative and equating to zero the derivatives over density function.
So, at least for this case no differences in results are possible. 
From our point of view it is much more reasonable to assume that in the 3DRISM model the pressure differs from the MDFT pressure.
This assumption seems even more reasonable, if we take into account that 3DRISM theory describes the mixture of sites connected in the complicated way, so it is completely possible that each site can give its own contribution to the pressure.
In such a way, we argue that the conclusions of Ref [Misin] do not contradict our results for the MDFT functional and only additionally indicate that the question of the applicability of the Pressure Correction to the 3DRISM calculations should be the topic of additional investigation.

\section{ Conclusion }

\subtopic{
We derived a thermodynamically consistant HNC+HSB. The pressure is correct, to the price of the loss of an empirical parameter (Wu's HSBradius to have good SFE) $=>$ Overshooted correction. We introduce a mysterious supplementary correction ;) that makes everything work so fine.}

In our paper we performed an efficiency test of different Classical DFT functionals for the Solvation Free Energy calculations.
We confirmed the known that that the HNC approximation gives a hugely overestimated predictions of the SFE of the small molecules.
Using the model system of hard sphere solvated in the water we demonstrated that HNC functional cannot even qualitatively correct predict the dependency of the SFE on the size of the Hard Sphere.
We have shown that for the large spheres the SFE calculated with the HNC is proportional to the volume, instead of the surface area of the sphere.
We suggested that this can be connected with the fact that HNC approximation incorrectly reproduce the pressure of the system and as a consequence of it the $-P\Delta V$ contribution to the total solvation energy.
We derived the expression for calculation a compressibility-route pressure in the HNC model, and have shown that for the SPCE water used in our calculations it is 11.5 KBar. Thus the $P\Delta V$ term is dominant in the error of SFE calculations.
To improve predictions of the solvation free energies we proposed to use an aposteriori pressure correction \eqref{eq:PC}. 
We showed that HNC functional with the Pressure Correction (HNC/PC) correctly reproduces qualitative behavior of the SFE of hard spheres in water.
However, this correction is unable to overcome all the defficiencies of the HNC model. 
In particular we demonstrated that the solvent structure near the hard sphere is not reproduced correctly which also should have effect on the SFE calculations.
Although the errors of the SFE predictions  of the small molecules in HNC/PC are much smaller than in HNC approximation, nevertheless they are still essentially large.
To improve the HNC/PC functional we used the Hard Sphere Bridge (HS-Bridge) functional \eqref{eq:FHS} which includes the many-body correlation functions of hard spheres derived from the FMT theory.
This functional depends on the parameter $R$ - hard sphere radius of the solvent molecules.
We parameterized the functional to reproduce correctly the SFE of methane in water.
We showed that using this parameterization allows one to reproduce correctly SFEs of the small molecules in water.
However, this functional incorrectly reproduces the solvent pressure and as a result gives an incorrect behavior for the SFEs of large hard spheres in water.
This functional is also unable to reproduce the distribution of the solvent molecules near the hard sphere ball.

We used an alternative approach and performed another parameterization of the HS-Bridge functional, to get the correct pressure of the system.
The functional parameterized in this way can much better reproduce solvation parameters of the hard sphere in water. Both: solvation free energy and solvent structure behavior are qualitatively correct. 
However in this case SFEs of the small molecules in water  are essentially underestimated. 
We additionally tested ``hybrid'' functionals, which include both: HS-Bridge and Pressure Corrected and showed that for any reasonable value of the $R$-parameter these functionals always underestimate the solvation free energy of small solutes.
In such a way we came to the conclusion that we need to look for a compromise between the thermodynamical consistency and relevant results for the model systems and accuracy of the SFE predictions for the ``real'' systems like hydrated organic molecules.
In addition, we studied dependency of the errors of the SFE predictions with the functionals which reproduce the correct pressure on the partial molar volume of the molecule, and noted that the errors  can be reasonably well corrected by adding additional $\rho kT \Delta V$ term. 
On the basis of these considerations we proposed an additional correction which includes that component.
To check whether this correction is universal or not we performed the SFE calculations for 122 organic molecules, and showed that for both studied functionals results with the additional correction are the best. In case of the HNC/PC+ functional the errors of predictions are comparable with the errors of simulations.
So, this allows us to suggest  that this correction should be used in the MDFT calculations to achieve the best accuracy of the SFE predictions.
The choice of the functional - HNC/PC+, HS/PRES+ or other depends on the particular task, and is the matter of question which of the thermodynamical properties are important in the particular research. 


\bibliographystyle{unsrt}
\bibliography{sergiievskyi_pmv_hs}

\end{document}